\documentclass[11pt]{article}
\usepackage{graphicx}
\usepackage{psfrag}
\usepackage{amssymb}
\begin{document}
\title{The Bowl Championship Series:\\ A Mathematical Review}
\author{Thomas Callaghan, Peter J. Mucha, and Mason A.
  Porter\footnote{Thomas Callaghan is an undergraduate majoring in
    applied mathematics, Peter Mucha is an assistant professor of
    mathematics, and Mason Porter is a VIGRE visiting assistant
    professor, all at Georgia Institute of Technology.  This work was
    partially supported by NSF VIGRE grant DMS-0135290, as a Research
    Experiences for Undergraduate project, and by a Georgia Tech
    Presidential Undergraduate Research Award. The simulated monkeys
    described herein do not know that they live on Georgia Tech
    computers. No actual monkeys were harmed in the course of this
    investigation.}}
\maketitle

\section*{Introduction}

On February 29, 2004, the college football Bowl Championship Series
(BCS) announced a proposal to add a fifth game to the ``BCS bowls'' 
to improve access for mid-major
teams ordinarily denied invitations to these lucrative
postseason games.  Although still subject to final
approval, this agreement is expected to be instituted
with the new BCS contract just prior to the 2006
season.

There aren't too many ways things could have gone worse this 
past college football season with the BCS Standings that 
govern which teams play in the coveted BCS bowls. The controversy
over USC's absence from the BCS National Championship game,
despite being \#1 in both polls, garnered most of the media attention
\cite{BCS-USC}, but it is the yearly treatment received by the
``non-BCS'' mid-major schools that appears to have finally
generated changes in the BCS system \cite{BCS-changes}.

Created from an abstruse combination of polls, computer rankings,
schedule strength, and quality wins, the BCS Standings
befuddle most fans and sportswriters, as we repeatedly get 
``national championship'' games between purported ``\#1'' 
and ``\#2'' teams in disagreement with the polls'
consensus. Meanwhile, the
top non-BCS squads have never been invited to a BCS bowl. Predictably,
some have placed blame for such predicaments squarely on the 
``computer nerds'' whose ranking algorithms form part of the BCS formula
\cite{nerds,nerds2}.  
Although we have no part in the BCS system, and the moniker 
may be accurate in our personal cases, we provide here a 
mathematically-inclined review of the BCS.  We briefly 
discuss its individual 
components, compare it with a simple algorithm defined by random 
walks on a biased graph, attempt to predict
whether the proposed changes will truly lead to increased BCS bowl access 
for non-BCS schools, and conclude by arguing that the true problem 
with the BCS Standings lies not in the computer algorithms, but rather in 
misguided addition.

\section*{Motivation for the BCS}

The National Collegiate Athletic Association (NCAA) neither conducts a
national championship in Division I-A football nor is directly
involved in the current selection process. For decades, teams were
selected for major bowl games according to traditional conference
pairings.  For example, the Rose Bowl featured the conference
champions from the Big Ten and Pac-10.  Consequently, a match between
the \#1 and \#2 teams in the nation rarely occurred.  This frequently left
multiple undefeated teams and co-champions---most
recently Michigan and Nebraska in 1997.  It was also possible for a
team with an easier schedule to go undefeated 
without having played a truly ``major'' opponent
and be declared champion by the polls, 
though the last two schools outside the current BCS agreement 
to do so were BYU in
1984 and Army in 1945.

The BCS agreement, forged between the six major ``BCS'' conferences (the
Pac-10, Big 12, Big Ten, ACC, SEC and Big East, plus
Notre Dame as an independent), was instituted in 1998 in an
attempt to fix such problems by matching the top two NCAA Division I-A
teams in an end-of-season BCS National Championship game. The BCS
Standings, tabulated by The National Football
Foundation \cite{NFFWeb}, selects the champions of the BCS conferences
plus two at-large teams to play in four end-of-season ``BCS bowl
games,'' with the top two teams playing in a National Championship
game that rotates among those bowls. Those four bowl
games---Fiesta, Orange, Rose, and Sugar---generate more than \$100
million annually for the six BCS conferences, but less than 10\% of
this windfall trickles down to the other five (non-BCS) Division I-A 
conferences \cite{BCS-AP}. With the current system only guaranteeing a
BCS bowl bid to a non-BCS school that finishes in the top 6 in the
Standings, those conferences have complained that their barrier to
appearing in a BCS bowl is unfairly high
\cite{NonBCScomplain}.  Moreover, the money directly generated by 
the BCS bowls is only one piece of the proverbial pie, as the schools 
that appear in such high-profile games receive marked increases in 
both donations and applications.

Born from a desire to avoid controversy, the short history
of the BCS has been anything but uncontroversial. In 2002, precisely two 
major teams (Miami and Ohio State) went undefeated during the regular season, 
so it was natural for them to play each other
for the championship. In 2000, 2001, and 2003, however, three or four teams 
each year were arguably worthy of claiming one of the
two invites to the championship game. Meanwhile, none of the non-BCS
schools have ever been invited to play in a BCS bowl. Tulane went
undefeated in 1998 but finished 10th in the BCS Standings. Similarly,
Marshall went undefeated in 1999 but finished 12th in the BCS. In
2003, with no undefeated teams and six one-loss teams, the three BCS
one-loss teams (Oklahoma, LSU, and USC) finished 1st through 3rd 
(respectively) in the BCS Standings, whereas the three non-BCS one-loss teams
 finished 11th (Miami of Ohio), 17th (Boise State), and 18th (TCU).

The fundamental difficulty in accurately ranking or even agreeing on a
system of ranking the Division I-A college football teams lies in two
factors---the paucity of games played by each team and the large
disparities in the strength of individual schedules.
With 117 Division I-A football teams, the 10--13 regular season
games (including conference tournaments) played by each team severely
limits the quantity of information relative to, for example, college
and professional basketball and baseball schedules.  While the 32 teams
in the professional National Football League (NFL) each play 16
regular season games against 13 distinct opponents,
the NFL subsequently uses regular
season outcomes to seed a 12-team playoff. Indeed, Division I-A
college football is one of the only levels of any sport that doesn't
currently determine its champion via a multi-game playoff
format.\footnote{The absence of a Division I-A playoff is itself quite
  controversial, but we do not intend to address this issue here;
  rather, we are more immediately interested in possible solutions
  under the constraint of the NCAA mandate against playoffs.}  Ranking 
teams is further complicated by the Division
I-A conference structure, as teams play most of their games within their
own conferences, which vary significantly in their level of play. 
To make matters worse, even the notion of ``top 2'' teams is woefully 
nebulous:  Should these be the two teams who had the best aggregate season 
or those playing best at the end of the season?

\section*{The BCS formula and its components}

In the past, national champions were selected by polls, which have been 
absorbed as one component of the BCS formula. However,
they have been accused of bias towards
the traditional football powers and of making only conservative changes
among teams that repeatedly win.  In attempts to provide unbiased
rankings, many different systems have been promoted by mathematically
and statistically inclined fans.  A subset of these algorithms comprise
the second component of the official BCS Standings.  Many of these schemes
are sufficiently complicated mathematically that it is virtually 
impossible for lay sports enthusiasts to understand them.  Worse still, 
the essential ingredients of some of the algorithms 
currently used by the BCS are not publicly declared. This state of 
affairs has inspired the creation of software to develop one's own 
rankings using a collection of polls and algorithms \cite{Rankulator} 
and comical commentary on ``faking'' one's own mathematical 
algorithm \cite{Kirlin}.

Let's break down the cause of all this confusion. The BCS Standings
are created from a sum of four numbers: polls, computer
rankings, a strength of schedule multiplier, and the number of losses by
each team. Bonus points for ``quality wins'' are also awarded for victories
against highly-ranked teams.  The smaller the resulting sum for a 
given team, the higher that team will be ranked in the BCS Standings.

The first number in the sum is the mean ranking earned by a team in the AP 
Sportswriters Poll and the USA Today / ESPN Coaches Poll.

The second factor is an average of computer rankings. Seven sources
currently provide the algorithms selected by
the BCS. The lowest computer ranking of each
team is removed, and the remaining six are averaged.  The sources of
the participating ranking systems have changed over the short history
of the system, most recently when the BCS mandated that the
official computer ranking algorithms were not
allowed to use margin of victory starting with the 2002 season. In the two 
seasons since that change, the seven official systems have
been provided by Anderson \& Hester, Billingsley, Colley, Massey, The
New York Times, Sagarin, and Wolfe. None of these sources receive any
compensation for their time and effort; indeed, many of them appear to
be motivated purely out of a combined love of football and
mathematics. Nevertheless, the creators of most of these systems guard
their intellectual property closely. An exception is Colley's ranking,
which is completely defined on his web site \cite{ColleyWeb}. 
Billingsley \cite{BillingsleyWeb}, Massey \cite{MasseyWeb}, and 
Wolfe \cite{WolfeWeb} provide significant information about the
ingredients for their rankings, but it is insufficient to 
reproduce their analysis.  Additional information about the BCS computer 
ranking algorithms (and numerous other ranking systems) can be found on 
David Wilson's web site \cite{dwilson}.

The third component of the BCS formula is a measurement of each team's 
schedule strength.  Specifically, the BCS uses a variation of what is 
commonly known in sports as the
Ratings Percentage Index (RPI), which is employed in college
basketball and college hockey to help seed their end-of-season
playoffs.  In the BCS, the average winning percentage of each team's
opponents is multiplied by 2/3 and added to 1/3 times the winning
percentage of its opponents' opponents. This schedule strength is
used to assign a rank to each team, with 1 assigned to that deemed 
most difficult.  That rank ordering is then 
divided by 25 to give the ``Schedule Rank,'' the third additive component of
the BCS formula.

The fourth additive factor of the BCS sum is the total number of losses by each
team. 

Once these four numbers (polls, computers, schedule strength, and losses) are 
summed, a final quantity for ``quality wins'' is
subtracted to account for victories against top teams. The current reward is
$-1.0$ points for beating the \#1 team, decreasing in magnitude in
steps of $0.1$, down to $-0.1$ points for beating the \#10 team.

It is not difficult to imagine that small changes in any of the above 
weightings have the potential to alter the BCS Standings dramatically.  
However, because of the large number of parameters, including unknown
`hidden parameters' in the minds of poll voters and the algorithms
of computers, any attempt to exhaustively survey possible changes 
to the rankings is hopeless. Instead, to 
demonstrate how weighting different factors can influence the rankings, 
we discuss a simple ranking algorithm in terms of random walkers on a 
biased network.

\section*{Ranking football teams with random walkers}

Before introducing yet another ranking algorithm, we emphasize that
numerous schemes are available for ranking teams in all sports.  
See, for example, \cite{Keener93,ConnorGrant00,Martinich02} for reviews of 
different ranking methodologies and the listing and bibliography maintained
online by David Wilson \cite{dwilson}.

Instead of attempting to incorporate every conceivable factor that 
might determine a team's quality, we took a minimalist approach,
questioning whether an exceptionally naive
algorithm can provide reasonable rankings.  We consider 
a collection of random walkers who can each cast a single vote for the 
team they believe is the best.  Their behavior is defined so
simplistically that it is reasonable to think of them as a large collection 
of trained monkeys.  Because the most natural arguments concerning the
relative ranking of two teams arise from the outcome of head-to-head
competition, each of these monkeys routinely examines the outcome of a
single game played by their favorite team---selected at random from that
team's schedule---and determines its new vote based entirely on the outcome 
of that game, preferring but not absolutely certain to go with the winner.

In the simplest definition of this process, the probability $p$ of
choosing the winner is the same for all voters and games
played, with $p > 1/2$ because on average the winner should be the
better team and $p < 1$ to allow a simulated monkey to
argue that the losing team is still the better team (due perhaps 
to weather, officiating, injuries, luck, or the phase of the
moon).  The behavior of each virtual monkey is driven by a simplified
version of the ``but my team beat your team'' arguments one
commonly hears. For example, much of the 2001 BCS controversy
centered on the fact that BCS \#2 Nebraska lost to BCS \#3
Colorado, and the 2000 BCS controversy was driven by BCS \#4
Washington's defeat of BCS \#3 Miami and Miami's win over BCS \#2
Florida State.

The synthetic monkeys act as independent random walkers on a
graph defined with biased edges
between teams that played head-to-head games.  This algorithm is easy
to define in terms of the microscopic behavior of individual
monkeys who randomly change their votes based on the win-loss outcomes
of individual games. The random behavior of these individual voters
is, of course, grossly simplistic. Indeed, under the specified range
of $p$, a given voter will never reach a certain conclusion about
which team is the best; rather, it will forever change its allegiance
from one team to another, ultimately traversing the entire graph. In
practice, however, the macroscopic total of votes cast for
each team by an aggregate of random-walking voters quickly reaches a
statistically-steady ranking of the top teams according to the quality
of their seasons.

We propose this model on the strength of its simple interpretation of
random walkers as a reasonable way to rank the top teams (or at
least as reasonable as other available methods, given the scarcity of
games played relative to the number of teams).  This simple scheme
has the additional advantage of having only one explicit,
precisely-defined parameter with a meaningful interpretation
easily understood at the level of single-voter behavior.  We have
investigated the historical performance and mathematical properties of
this ranking system elsewhere \cite{MonkeyManuscript,MonkeyWeb}.
At $p$ close to $1/2$, the ranking is dominated by an RPI-like ranking 
in terms of a team's record, opponent's records, etc., with little 
regard for individual game outcomes. For $p$ near $1$, on the other hand, 
the ranking depends strongly on which teams won and lost against 
which other teams.

Our initial questions can now be rephrased playfully as follows: Can a bunch 
of monkeys rank football teams as well as the systems currently in
use?  Now that we have crossed over into the Year of the Monkey in
the Chinese calender and the BCS has recently proposed changes 
to their non-BCS rules, it seems reasonable to ask whether the monkeys can
clarify the effects of these planned changes.

\section*{Impact of proposed changes on non-BCS schools}

The complete details of the new agreement have not yet been released, 
but indications are that the proposed rules would have
given four at-large BCS bids to non-BCS schools over the past six
years \cite{BCS-AP}. Based on the BCS Standings, the best
guesses at those four teams are 1998 Tulane (11-0, BCS \#10, poll
average 10), 1999 Marshall (12-0, BCS \#12, poll average 11), 2000 TCU
(10-1, BCS \#14, poll average 14.5), and 2003 Miami of Ohio (12-1, BCS
\#11, poll average 14.5). However, there are also indications that
only non-BCS teams finishing in the BCS top 12 would automatically get
bids \cite{BCS-changes}, and each of these four schools would have had to be 
given one of the at-large bids over at least one team ahead of them 
in the BCS Standings \cite{BCS-silly}.

Given the perception that the polls unfairly favor BCS schools, it
is worth noting the contrary evidence from six seasons of BCS
Standings. In addition to the four schools listed above, other notable
non-BCS campaigns were conducted this past season by Boise State (12-1,
BCS \#17, poll average 17) and TCU (11-1, BCS \#18, poll average 19).
Five of these six schools earned roughly the same ranking in the BCS 
standings and the polls.  The only significant exception was 2003 
Miami of Ohio, averaging 6th in the official BCS computer 
algorithms but only 14.5 in the polls.

While the new rules might indeed give BCS bowl bids to all non-BCS schools
who finish in the top 12, it is worth inquiring how
close non-BCS schools may have come to this or 
to a top 6 ranking that would have
guaranteed them a bid during the past six years. In particular, 2003 was 
the first time in the BCS era that there were no undefeated
teams remaining prior to the bowl games.  Given that there were six
one-loss teams and no undefeateds, what would have happened if one or
more of the three non-BCS teams had instead gone undefeated? While it is
impossible to guess how the polls would have behaved and we are
unable to reproduce most of the official computer rankings, we can instead
compute the resulting ``random-walking monkey'' rankings for different
values of the bias parameter $p$. As a baseline, Figure 1 plots 
the end-of-season, pre-bowl-game rankings of each of 
the six one-loss teams, plus Michigan, from the true 2003 season 
(scaled logarithmically so that the top 2, top 6, and top 12 teams are 
clearly designated).

\begin{figure}[htbp]
  \centering
  {\psfragscanon\large
    \psfrag{Rank}[t][t]{Rank}
    \psfrag{p}[b][t]{$p$}
    \includegraphics[width=3.0in]{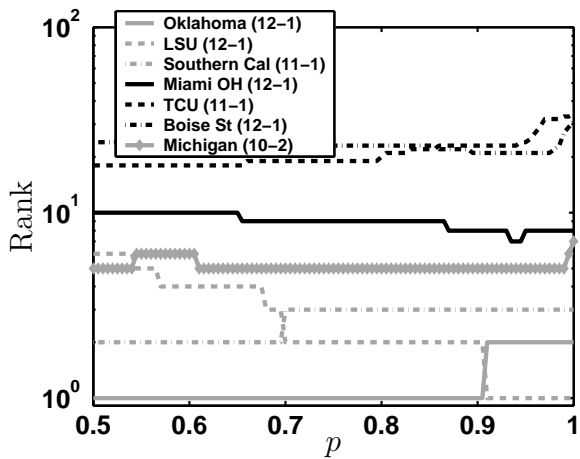}
  }
  \caption{Random-walking monkey rankings of selected teams for 2003.}
  \label{fig:1}
\end{figure}

Now consider what would have transpired had Miami of Ohio, TCU, and
Boise State all gone undefeated. Figure 2 shows the resulting rankings
of the same teams as Figure 1 under these alternative outcomes.
In the limit $p\to 1$, going undefeated trumps
any of the one-loss teams, so each of these mythically undefeated
schools ranks in the top 3 in this limit. For TCU and Boise State,
however, their range of $p$ in the top 6 is quite
narrow. If the new rules require only a top 12 finish for a non-BCS
team, then the situation looks much brighter for an undefeated TCU, which 
earned monkey rankings in the top 11 at all $p$ values. However, according 
to the scenario plotted in Figure 2, an undefeated Boise State's claim on 
a BCS bid remains tenuous even under the proposed changes.
Indeed, even had Boise State been the only undefeated team last
season (not shown), the monkeys would have left them out of the top 10 and 
behind Miami of Ohio for $p \lesssim 0.86$.

\begin{figure}[htbp]
  \centering
  {\psfragscanon\large
    \psfrag{p}[b][t]{$p$}
    \psfrag{Rank}[t][t]{Rank}
    \includegraphics[width=3.0in]{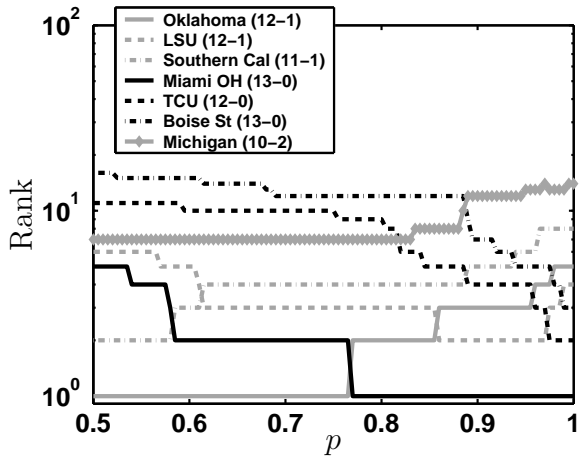}
  }
  \caption{Random-walking monkey rankings of selected teams for an 
    ``alternate universe'' 2003 in which the three non-BCS, one-loss teams
    instead went undefeated.}
  \label{fig:2}
\end{figure}

At the other extreme, one-loss Miami of Ohio already has a legitimate
claim to the top 12 according to both the monkeys \emph{and} the real BCS
Standings.  Note, in particular, the exalted ranking the monkeys would have
given Miami of Ohio had they won their season opener against
Iowa (their only loss in the actual 2003 season). According to the
monkeys, they may have even had a reasonable argument to be placed in
the championship game had they gone undefeated. It was bad
enough not being able to fit 3 teams onto the field for the BCS National
Championship game, but we might have been one Miami of Ohio
victory over Iowa away from wanting to crowd four squads into the mix!

As an example of how the effects of games propagate into the rankings
of other teams, we also include Michigan's ranking in both figures, even 
though their outcomes were not changed in the calculations that produced 
the two plots.  Nevertheless, because Michigan is a next-nearest neighbor 
of Miami of Ohio in the network (both teams lost to Iowa in 2003), 
changing the outcome of the Iowa v.\ Miami of Ohio game unsurprisingly 
affects Michigan's ranking detrimentally.

To conclude this section, we stress that the above discussion is purely 
hypothetical, as the monkeys only provide a stand-in for our inability to 
compute true BCS Standings under alternative outcomes.

\section*{The problem at the top, and a possible solution}

While we focused above on non-BCS schools and the recent
changes that improve their chances of playing in a BCS bowl game,
the larger BCS controversy for many fans is the recurring
inability of the BCS to generate a championship game between
conclusive ``top 2'' teams. Each of the past four
seasons, the two polls agreed on the top 2 teams prior to the bowl
games.  In three of those seasons, however, the top 2 spots in the BCS
Standings included only one of the teams selected by the polls.  
In 2000 and 2001, the \#2 team in the polls ended up on the short 
end of the BCS stick, whereas in 2003 it was USC (the \#1 team in both polls)
on the outside looking in. 

Although it is easy to blame this situation on the computer rankings,
the true problem as we see it lies in the BCS formula of polls,
computers, schedule strength, losses, and quality wins.  Simply,
the polls and computers already account for schedule
strength and ``quality wins'' or else the three non-BCS one-loss
teams (Miami of Ohio, TCU, and Boise State) would have placed in the top
6 in the 2003 BCS Standings.  Adding these factors \emph{again} after
the polls and computer rankings are determined disastrously double-counts 
these effects, adversely degrading confidence in the BCS selections for
the National Championship and the other BCS bowls.

One of the presumed motivations for including separate factors
for schedule strengh and quality wins was to reduce the
assumed bias of the polls towards traditional football powers.
However, as discussed above, the top non-BCS teams over the past six
years were ranked similarly in the polls and computers.
Therefore, one might rightly worry that the quality wins and 
schedule strength factors are making it harder for non-BCS
schools to do well in the standings, as their schedules are
typically ranked significantly lower and they have few
opportunities for so-called ``quality wins.''

USC was on the losing end of this double counting in 2003, having finished the
regular season \#1 in both polls and averaged \#2.67 on the computers. LSU 
was \#2 in both polls and averaged \#1.93 on the
computers, and Oklahoma was \#3 in both polls and averaged \#1.17 on the
computers. One of the official computer systems even ranked non-BCS Miami of
Ohio ahead of USC. However, although the computers ranked Oklahoma ahead
of the other teams, it was Oklahoma's 11th place schedule
strength and $-0.5$ ``quality win'' bonus for beating Texas that combined 
to give it an
additional 1.55 BCS points edge compared to USC's 37th place schedule
(standings available from \cite{NFFWeb}).
With six one-loss
teams in Division I-A, the ranking algorithms predominantly
favored Oklahoma \emph{because} of its relatively difficult schedule
and its victory over Texas. Without those effects being included
\emph{again} in separate quality wins and schedule strength factors, a
straight-up averaging of the polls and the computers would rank USC 
first (1+2.67=3.67), LSU second (2+1.93=3.93), 
and Oklahoma third (3+1.17=4.17).

A reasonable knee-jerk reaction to this proposal would be to reassert
that schedule strength, number of losses, and so-called quality wins
should matter. Our point is that they are \emph{already} incorporated in such 
a simple averaging scheme, as the polls and the computers (necessarily) 
consider such factors to produce reasonable rankings.
To explicitly add further BCS points for each of these considerations 
gives them more weight than the collective wisdom of the polls and 
computer rankings believe they should have.

Whatever solution is ultimately adopted, we strongly advocate that
modifications to the BCS remove such double-counting and,
ideally, provide a system that is more open to the community. That the
double-counting problem isn't widely appreciated further supports our
opinion that the BCS system needs to be more transparent.  {\it
  The recently announced changes do not address this problem.}
College football fans shouldn't have to accept computer rankings
without a minimal explanation of their determining ingredients,
not only so that they have more confidence in these algorithms
but also to open debate about what factors should be included and 
how much they should be weighted.
For example, there is certainly a need
to discuss how much losing a game late in the season or in a
conference championship game (as Oklahoma did in 2003) should matter
compared to an earlier loss.  

Even before the end-of-season
controversy in 2003, a survey conducted by New Media
Strategies indicated that 75\% of college football fans thought that
the BCS system should be scrapped entirely \cite{BCS-survey}.  That number 
presumably increased after the new round of controversy.
Changes that lead to greater transparency and a simplified weighted
averaging of the polls and computers are the only way anything
resembling the current BCS system can maintain popular support.


\end{document}